Research Article

# Structuring Polarization States of Light in Space and Time

Danilo Gomes Pires[†], Jiaren Tan[†], Hooman Barati Sedeh and Natalia M. Litchinitser[*]

**Abstract:** The spatiotemporal sculpturing of light beams with arbitrary phase and polarization topologies has garnered significant attention in recent years due to its potential to advance optical technologies and reveal novel physical phenomena. Examples of spatiotemporal beams include space–time wave packets, flying donuts, tilted pulse fronts, X-waves, Airy pulses, and spatiotemporal optical vortices. Here, we introduce and demonstrate a new class of spatiotemporal polarization states of light. We propose a generalized spatiotemporal higher-order Poincaré sphere and show that these polarization states emerge from the superposition of two orthogonal circular polarization states, each carrying a spatiotemporal optical vortex. Such a choice of the basis enables simultaneous control of the spatial and temporal degrees of freedom of light. Theoretical predictions are experimentally validated using ultrafast femtosecond pulses, revealing how the resulting polarization distributions evolve in both space and time. Finally, we further extend this approach to construct a family of spatiotemporal skyrmionic textures that are localized, topologically nontrivial configurations of the electromagnetic field vector, offering a versatile framework for generating and controlling multidimensional (space and time) structured polarization fields. The ability to create and manipulate diverse forms of spatiotemporal skyrmionic textures opens up new opportunities for studying complex light-matter interaction phenomena, advanced imaging and micromanipulation, and encoding information across both space and time, with potential implications for advanced optical communication and information processing in classical and quantum domains.

**Corresponding author**: **Natalia M. Litchinitser**, Department of Electrical and Computer Engineering, Duke University, Durham, NC, USA. E-mail: natalia.litchinitser@duke.edu

**Danilo Gomes Pires, Jiaren Tan, and Hooman Barati Sedeh**, Department of Electrical and Computer Engineering, Duke University, Durham, NC, USA.

[†]These authors contributed equally to this work

## 1. Introduction

Conventional optics has traditionally emphasized plane waves and Gaussian beams; however, advances in singular optics and the orbital angular momentum (OAM) of light have introduced the concept of structured light. This development has expanded the capabilities of modern optics by enabling the generation and control of complex light fields with customized spatial and phase characteristics, opening new avenues for research and applications in imaging, communications, and quantum science. Spatially structured light beams include optical waveforms with spatial inhomogeneities in two or three dimensions, including beams with complex spin angular momentum (SAM) and OAM in both paraxial and non-paraxial regimes [1-4], and vector beams with inhomogeneous polarization distributions [5-7]. Structured vector beams can be mathematically represented using the Poincaré sphere (PS) [8] and the higher-order Poincaré sphere (HOPS) [9], and find applications in particle micromanipulation [10, 11], optical communications [12, 13], and sensing [14, 15].

In addition to spatial control, recent advances in temporal shaping of optical beams have introduced a new degree of freedom, enabling four-dimensional (x, y, z, t) manipulation of light [16]. The combination of spatial and temporal beam shaping has given rise to spatiotemporal structured beams, such as spatiotemporal optical vortices (STOVs) [17-19] that possess both transverse and longitudinal OAM components. While the original spatiotemporal optical vortices (STOVs) were constructed from scalar fields, they did not account for the vectorial nature of light. In this work, we introduce the spatiotemporal higher-order Poincaré sphere (STHOPS) as a novel framework for representing polarization dynamics in space and time. This approach extends the concept of STOVs beyond scalar field distributions, enabling a unified description of spatial, temporal, and polarization degrees of freedom within a single representation. By modulating orthogonal vector functions across both space and time, we achieve continuous control over the polarization states on STHOPS, enabling the generation of a wide variety of spatiotemporal polarization vortices. While a spatio-spectral version of the Poincaré sphere was recently proposed to describe polarization across the frequency components of a pulse [24], our approach directly addresses the combined spatial and temporal domains. We further investigate how spectral transformations affect beam behavior, uncovering distinct polarization distributions in the spectral domain and unique polarization states in the temporal domain. Finally, we utilize this platform to create spatiotemporal optical skyrmionic textures, which represent topologically nontrivial polarization configurations that evolve in both space and time. These textures not only provide a new route for exploring fundamental aspects of light's topology but also open exciting possibilities for various applications such as high-capacity information encoding, ultrafast optical communication, and advanced signal processing.

## 2. Theoretical model

The polarization states of light are commonly mapped into the well-known PS, where each unique point on its surface represents a particular state of polarization [8]. Within the spatial domain, the homogeneous polarization states of light can be described by a superposition of two orthogonal circular polarizations as

$$|E_{\text{PS}}\rangle = \cos\left(\frac{\theta}{2}\right)|\sigma^+\rangle + \sin\left(\frac{\theta}{2}\right)e^{i\phi}|\sigma^-\rangle, \quad (1)$$

where $|\sigma^+\rangle = \vec{x} + i\vec{y}$ and $|\sigma^-\rangle = \vec{x} - i\vec{y}$ correspond to the right (RCP) and left (LCP) circular polarizations, and $\vec{x}$ and $\vec{y}$ are the orthogonal linear polarizations along the horizontal and vertical directions, respectively; the azimuthal ($\phi$) and the polar ($\theta$) angles determine the orientation and ellipticity of the polarization states which can be retrieved in the experiment by measuring the Stokes parameters [20]. We note that while Eq. (1) enables the generation of different homogeneous polarization states along the PS, more complex, inhomogeneous polarization states can be achieved by linearly superposing two orthogonal circular polarizations carrying spatially structured amplitude and wavefront [9]. For instance, a higher-order PS (HOPS) [11] gives rise to a family of vectorial light fields, particularly those with spatially varying polarization. Taking as a basis the Laguerre-Gaussian (LG) modes, which consist of an orthogonal set of solutions of the paraxial wave equation in cylindrical coordinates and can be written at the waist plane $z = 0$ in polar coordinates $(r, \varphi)$ as [4, 21]

$$\mathrm{LG}_{p,l}(r,\varphi) = \sqrt{\frac{p!}{\pi(|l|+p)!}} \frac{r^{|l|}}{w_0^{|l|+1}} L_p^{|l|}\left(\frac{r^2}{w_0^2}\right) e^{-\frac{r^2}{2w_0^2}} e^{il\varphi}, \quad (2)$$

where the indices $(p, l)$ represent the radial and azimuthal indices, respectively, $w_0$ is the beam waist, and $L_p^{|l|}$ is the generalized Laguerre polynomials with the radial index $p$, and azimuthal index $l$, which is known as topological charge, characterizes the phase singularity at the center of the beam. Eq. (1) can be rewritten as [9]

$$|E_{\mathrm{HOPS}}\rangle = \cos\left(\frac{\theta}{2}\right) \mathrm{LG}_{-l} |\sigma^+\rangle + \sin\left(\frac{\theta}{2}\right) \mathrm{LG}_l e^{i\phi} |\sigma^-\rangle, \quad (3)$$

where the optical modes $\mathrm{LG}_{\pm l} = \mathrm{LG}_{p=0,\pm l}$ were introduced. Equation (3) describes the family of vectorial light states on the HOPS.. We note that Eq. (3) reduces to Eq. (1) when the constituent LG modes carry topological charge of $l = 0$.

Now, we introduce a method to construct an STHOPS by superposing two STOV states [18, 19] instead of the spatial-only LG beams. The spatiotemporal version of Eq. (3) is given by

$$\begin{aligned} |E_{\mathrm{STHOPS}}\rangle &= \cos\left(\frac{\theta}{2}\right) u(x,y,z,t)_{-l} |\sigma^+\rangle \\ &+ \sin\left(\frac{\theta}{2}\right) u(x,y,z,t)_l e^{i\phi} |\sigma^-\rangle \end{aligned} \quad (4)$$

where $\theta, \phi$ denote the same polar and azimuthal angles as in the case of the original PS, and

$$u(x, y, z = 0, t)_l = \sqrt{\frac{2}{\left(\frac{x}{w_0}\right)^2 + \left(\frac{t}{\tau}\right)^2}} \left(\frac{t}{\tau} + \mathrm{sgn}(l) \frac{ix}{w_0}\right)^{|l|} e^{-\frac{x^2+y^2}{w_0^2}} e^{-\frac{t^2}{\tau^2}} \quad (5)$$

stands for a $l$-th order STOV at $z = 0$, with $\tau$ denoting the pulse duration [18]. In contrast to the LG modes, STOVs possess a phase singularity line in the space-time domain, for example, in the direction perpendicular to the spatiotemporal plane $x,t$, and a spatial Gaussian envelope along the $y$-direction. Several techniques have been developed to generate spatiotemporal optical vortices (STOVs) in laboratory settings, including the use of pulse shapers combined with holographic devices [23–25] and optical metasurfaces [26–28]. Notably, the first observation of these states was made during filamentation in air [17, 22]. Figure 1 (a,b) shows the intensity and phase distributions of a −1 and +1 order STOV, respectively. An iso-intensity distribution is displayed in Fig. 1 (c) to illustrate the spatiotemporal profile of the STOV with order ±1.

## 3. Experimental realization of STHOPS states

Figure 2 shows an experimental setup developed to generate superposed spatiotemporal beams and perform a proper diagnostic within space, time, and polarization. A Ti: Sapphire pulsed laser with a pulse duration of 100 fs and a repetition rate of 1 kHz passes through an optical parametric amplifier, generating near-infrared pulses centered at 1550 nm. The initial beam is then separated using a 50/50 beam splitter (BS) into two routes, forming the signal and reference pulses. We rotate the polarization of the signal beam to 45° using a half-wave plate (HWP), followed by a polarized beam splitter (PBS) to further separate orthogonal linearly polarized pulses. Next, each of these orthogonally polarized pulses is redirected to the grating (G) and the output frequencies are collimated by a cylindrical lens (CL) before impinging on a spatial light modulator (SLM). Since the SLM only modulates light in one specific polarization state, an additional HWP has been introduced in one of the orthogonally polarized beams to ensure correct phase modulation. Spiral phase masks are used as phase-only holograms carrying opposite topological charges $l = \pm 1$ were spatially separated within the SLM window. After the spatiospectral modulation, the pulses travel back and are combined after the PBS. The spatiotemporal vector beam, consisting of opposite topological charge LG beams at right- and left-circular polarization states, as described by Eq. (3), forms after the quarter-wave plate (QWP). The beam structure is retrieved by combining Stokes polarimetry and a three-dimensional diagnostic technique [23, 24]. The linear components of the Stokes parameters are measured by adding a linear polarizer (LP) immediately before the InGaAs camera and rotating it by 0°, 90°, 45°, and 135°. Removing the linear polarizer allows us to measure the circular polarization components. The temporal profile of each spatiotemporal Stokes distribution is measured by scanning the modulated pulse with the reference beam, where the time delay between them is controlled by a delay stage (See Section 1 of the Supplementary Materials for the full description of the Stokes parameters). Note that the reference beam is configured to match the polarization state of the signal beam using an additional half-wave plate (HWP), which is essential for accurate interferometric measurements. The retrieved fields are then averaged over 10 measurements, with 100 temporal slices for each realization (For more details on the experimental averaging technique, see Section 2 of the Supplementary Materials).

The Stokes parameters $S_1, S_2, S_3$ indicate the polarization difference between the horizontal and vertical ($S_1$), anti-diagonal and diagonal ($S_2$), and right-hand and left-hand circular ($S_3$) states, respectively. In this context, the polarization difference between the states occurs within the spatiotemporal fields in Eq. (4), meaning that at each instant of time within the pulse, the beam possesses a different polarization distribution. In Fig. 3, we show several examples of the states within the +1-order STHOPS, described by intensity and polarization distributions at multiple $\theta, \phi$ angular locations along the +1-STHOPS surface, obtained from both theory and experiments. This is achieved by using Eq. (4) for the case of $l = 1$. Here, red circles denote RCP states ($|\sigma^+\rangle$) and blue circles stand for LCP states ($|\sigma^-\rangle$). We note that changing the topological charge to $l = -1$, we obtain another family of −1-order STHOPS, where several examples of the states are shown in Fig. 4. It should be noted that while Eq. (4) describes a higher-order polarization state in the spatiotemporal plane, its spatiospectral counterpart $(x, \omega)$ carries a different polarization distribution as it is outlined in Section 3 of the Supplementary Materials. An example of the Stokes parameters for the radially polarized state is shown in Fig. 5. The theoretical (a-d) and experimental (e-h) Stokes parameters $S_0, S_1, S_2, S_3$ in Fig. 5 correspond to the case of $l=1, \phi=0$, and $\theta=\pi/2$, along with their respective polarization distribution superposed in panels (a,e). Note that, due to the uneven intensity distribution in the experiments and the temporal mismatch of the superposing beams, some circular polarization leakage is observed. Despite these experimental imperfections, the polarization distribution from the STHOPS in Figs. 3 and 4 agree well with their theoretical counterparts.

## 4. Spatiotemporal skyrmionics textures

As described in the previous sections, the construction of STHOPS establishes a unified formalism for describing light fields whose polarization is both spatially structured and intrinsically coupled to their temporal evolution. In this representation, each point on the sphere corresponds to a distinct non-separable polarization-time mode, extending the concept into the spatiotemporal domain. Building on the theoretical framework established in Section 2, the constituent fields in the superposition (Eq. (4)) can be chosen to generate more complex spatiotemporal structures, allowing access to a broader range of polarization and topologically structured light fields. Let's now replace $u(x,y,z,t)_{-l}$ describing $-l$-order STOVs, by a Gaussian beam $G(x,y,t) = e^{-\frac{x^2+y^2}{w_0^2}} e^{-\frac{t^2}{\tau^2}}$, while keeping $u(x,y,z,t)_{-l}$ in the second term of Eq.(4) unchanged. This construction is given by

$$|E_S\rangle = \cos\left(\frac{\theta}{2}\right) G(x,y,t)|\sigma^+\rangle + \sin\left(\frac{\theta}{2}\right) u(x,y,z,t)_l e^{i\phi} |\sigma^-\rangle, \quad (6)$$

results in a formation of skyrmionic beams [31], which arise as topologically protected textures of the electromagnetic field, where the polarization vectors (or equivalently the local Stokes vectors) wrap the sphere in a nontrivial manner [34–38]. While these textures have been widely explored in the spatial domain, here we, for the first time, show the analogous spatiotemporal skyrmionic textures that are realized by appropriately tailoring the superposed fields, and extending the concept of optical skyrmions into the joint space–time domain. These structures exhibit polarization vortices and domain-wall-like features in combined space–time coordinates, enriching the accessible topological phase space.

Examples of the theoretical and experimental spatiotemporal textures corresponding to Néel-type optical skyrmions are shown in Fig. 6 (a) and 6(d), respectively, using $l=1, \theta=\pi/2, \phi=\pi$. The Bloch-type spatiotemporal optical skyrmions are also shown in Fig. 6 (b) and 6(e), with $l=1, \theta=\pi/2, \phi=3\pi/2$, respectively. Finally, the Anti-skyrmion-type spatiotemporal textures are shown in Fig. 6 (c) and 6(f) by using $l=-1, \theta=\pi/2, \phi=\pi$, respectively. In all cases, we can generate a given skyrmionic texture confined in the spatiotemporal domain.

The skyrmion numbers of the respective skyrmionic textures observed experimentally in Figure 6 (e-f) are calculated with respect to the reduced Stokes parameters $\mathbf{s} = (s_1, s_2, s_3)$ through [25]

$$N = \frac{1}{4\pi} \iint \mathbf{s} \cdot \left(\frac{\partial \mathbf{s}}{\partial x} \times \frac{\partial \mathbf{s}}{\partial \tau}\right) dx d\tau, \quad (7)$$

where each reduced Stokes parameters are written as $s_i = S_i/S_0$ and the integration is performed across the spatiotemporal cross-section, defined in the experiment as the spatial axis $x$ and the time delay between the signal and reference pulses $\tau$. The skyrmion numbers obtained from the experiments are found to be 0.95 for the Néel-type texture, 0.99 for the Bloch-type texture, and $-0.90$ for the Anti-skyrmion-type texture. Such skyrmionic textures are likely to enable new potential approaches for data storage and communication protocols, as well as optical sensing and metrology techniques.

In our experiments, for both STHOPS and spatiotemporal skyrmionic textures, the results were obtained using a phase-only hologram of the phase singularity in the (x, ω) plane, rather than a full amplitude and phase modulation of the Laguerre–Gaussian mode across the spatiotemporal domain. This technique is commonly used to generate optical vortices carrying orbital angular momentum without any radial orders at the far-field and was widely used to do so in both space-only [26-28] and spatiotemporal domains [18, 19, 23]. To validate this approach, Section 4 of the Supplementary Materials demonstrates that employing a phase-only hologram, as in the experimental arrangement, yields equivalent results.

## 5. Conclusion

In summary, we introduced spatiotemporal polarization distributions based on the superposition of STOVs possessing opposite phase dislocations on each orthogonal circular polarization state. The resulting families of states are determined by the STHOPS, controlled by adjusting the amplitude and phase difference between the corresponding field superposition and determining their Stokes parameters. Furthermore, upon varying the STOVs' order of each mode within the field superposition, we demonstrated novel spatiotemporal optical skyrmionic textures.

These new spatiotemporal polarization optical states are based on the complex interaction of polarization with spatial and temporal degrees of freedom, extending beyond traditional scalar or purely spatial vector beams. From a fundamental science standpoint, they enable studies of complex topological phase singularities and non-separable mode structures with potential implications for structured light-matter interactions, nonlinear optics, and ultrafast dynamics. These states are likely to provide multidimensional encoding capabilities for enhanced communication bandwidth, improved spatiotemporal resolution in imaging, and precise control in ultrafast laser systems. Their inherent high dimensionality is expected to enable advances in quantum information processing, positioning spatiotemporal polarization states as a promising platform bridging fundamental photonics and emerging technologies.

Acknowledgement- NML would like to acknowledge the support of this research by the Army Research Office Award no. W911NF2310057.

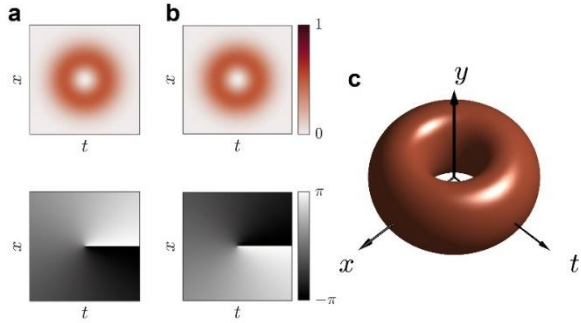

Fig. 1. Intensity and phase distributions for a STOV carrying spatiotemporal topological charge of (a) $l = -1$ and (b) $l = +1$. (c) Iso-intensity distribution of a STOV in both space and time domains.

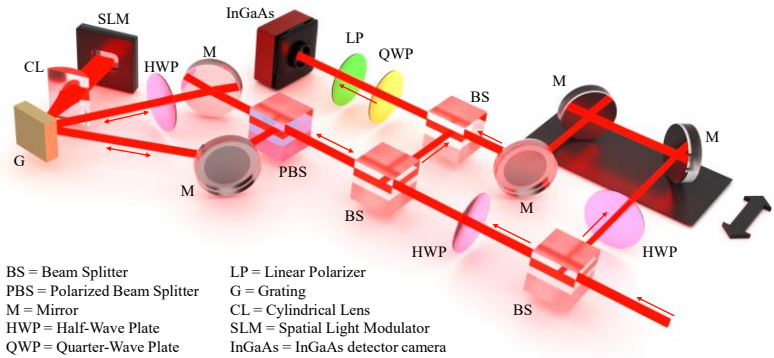

Fig. 2. Experimental setup for measuring spatiotemporal polarization distributions, which combines Stokes polarimetry and a three-dimensional diagnostics technique. The pulse shaper is formed by a grating with 600/mm grooves, a cylindrical lens with $f = 100$ mm, and a Hamamatsu LCOS-SLM X10468-08. The data is acquired with a Xenics Bobcat-320-GigE-8315 camera, possessing an InGaAs detector sensor. For more details on Stokes polarimetry, refer to the Supplemental Document.

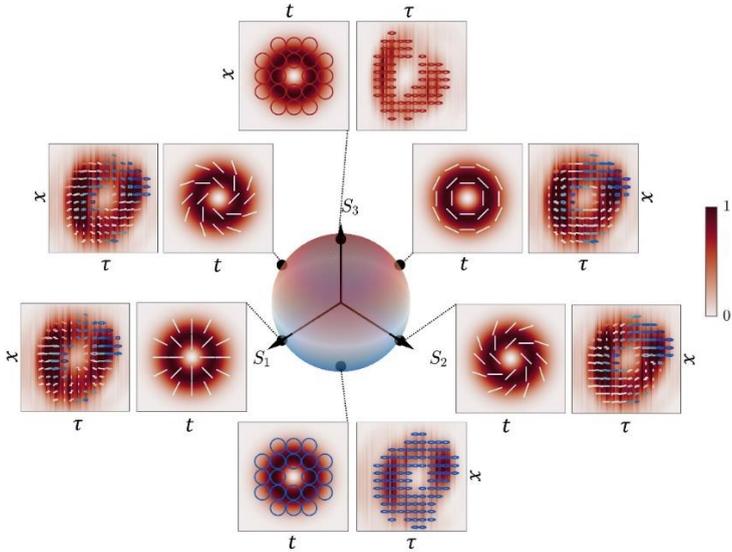

Fig. 3. Intensity and polarization distributions of the spatiotemporal vectorial states for the $+1$ order STHOPS. The North (South) pole corresponds to the RCP (LCP), and the states along the equator, $\theta = \pi/2$, correspond to $\phi = 0, \pi/2, \pi,$ and $3\pi/2$ with respect to $S_1$.

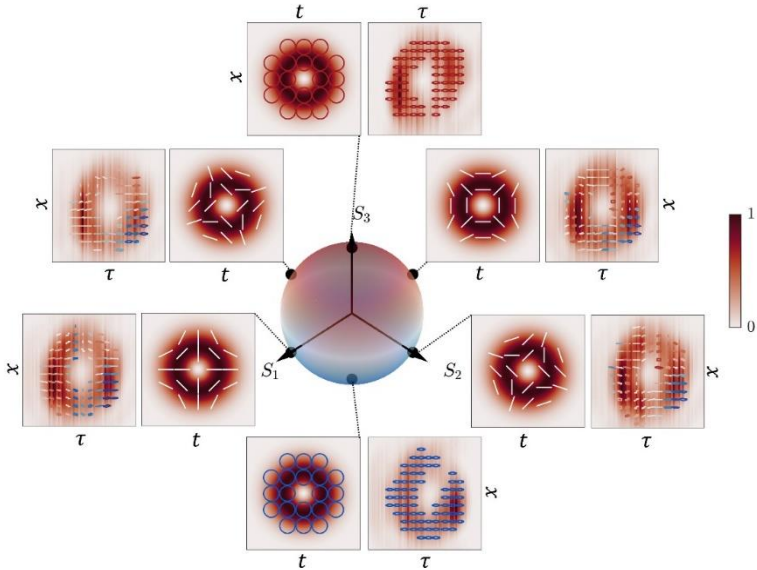

Fig. 4. Intensity and polarization distributions of the spatiotemporal vectorial states for the $-1$ order STHOPS. The North (South) pole corresponds to the RCP (LCP), and the states along the equator, $\theta = \pi/2$, correspond to $\phi = 0, \pi/2, \pi,$ and $3\pi/2$ with respect to $S_1$.

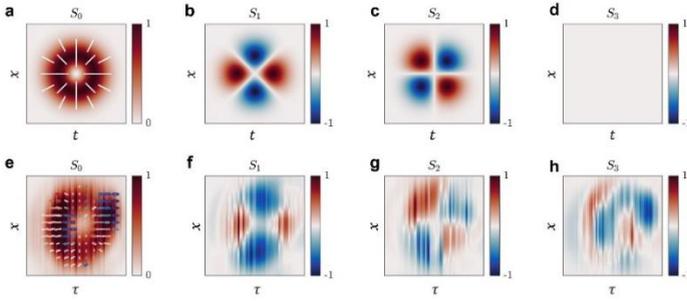

Fig. 5. Stokes parameters $(S_0, S_1, S_2, S_3)$ obtained from (a-d) theory and (e-h) experiments. Panels (a,e) show the theoretical and experimental intensity distribution together with their polarization states, respectively.

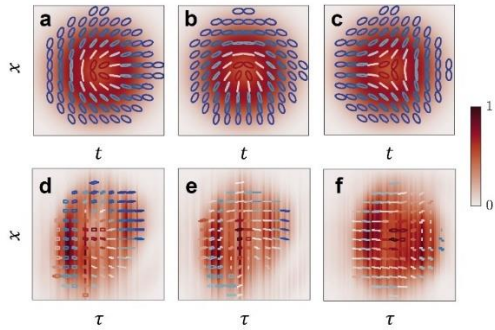

Fig. 6. Spatiotemporal skyrmionic textures constructed from (a-c) theory and (d-f) experiments. Here, we present the (a,d) Néel, (b,e) Bloch, and (c,f) Anti-skyrmion textures.